\documentclass[multphys,vecarrow]{svmult}

\usepackage{graphicx}
\usepackage{amssymb}
\usepackage{amsmath}
\usepackage{float}
\usepackage{subeqnar}
\usepackage{epsfig}
 
\begin{document}
\frontmatter
\mainmatter

\bibliographystyle{unsrt}

\title*{A Modification of the Social Force Model by Foresight\\
Preprint, to appear in the Proceedings of PED2008 }
\titlerunning{Modification by Foresight}
\author{Bernhard Steffen}
\institute{
Juelich Institute for Supercomputing, Forschungszentrum J\"ulich GmbH\\
52425 J\"ulich, Germany\\
{\tt b.steffen@fz-juelich.de}\\
}
\authorrunning{B. Steffen}

\date{\today}

\maketitle

\begin{abstract}
The motion of pedestrian crowds (e.g. for simulation of an evacuation situation)
can be modeled as a multi-body system of self driven particles 
with repulsive interaction \cite{HELB95,HELB00b,HELB03,SEY05a}. 
We use a few simple situations to determine the simplest allowed functional 
form of the force function. More complexity may be necessary to model more 
complex situations. There are many unknown 
parameters to such models, which have to be adjusted correctly 
to give proper predictions 
of evacuation times, local densities and forces on rails or obstacles. \\
The parameters of the social force model  \cite{HELB95,HELB00b,HELB03}
can be related to quantities that can be measured independently, like step length 
and frequency. The microscopic behavior is, however, only poorly reproduced in 
many situations, a person approaching  a standing or slow obstacle will e.g. show 
oscillations in position \cite{SEY05a}, 
and the trajectories of two persons meeting in a corridor 
in opposite direction  will be far from realistic and somewhat erratic. \\
One of the reasons why these models are not realistic is the assumption of 
instantaneous reaction on the momentary situation. Obviously, persons react with a 
small time lag \cite{HOOG05b}, while on the other hand they will anticipate changing situations for at
least a short time. Thus basing the repulsive interaction not on the 
momentary situation but on a (linear) extrapolation over a short time (e.g. 1 s) 
eliminates the oscillations at slowing down and smoothes the patterns of giving way 
to others to a more realistic behavior. The exact extrapolation time is of 
little importance, but a combination of long time with linear extrapolation may get
unstable. One second anticipation seems reasonable, and while the actual 
anticipation in peoples mind will most likely not be based on linear extrapolation, the 
differences will be small. \\
A second reason is the additive combination of binary interactions. It is shown that 
combining only a few relevant interactions gives better model performance. 

\noindent
\end{abstract}

The motion of pedestrians can be modeled as self driven particles.
This requires the definition of an accelerating (or  decelerating) force
depending on the situation - parameters of the pedestrian (walking speed, 
fitness, motivation),
the next (intermediate) goal (door etc), the restrictions placed by obstacles,
and position and movement of other persons in the neighborhood.
The best known such model is the social force model \cite{HELB95}
which assumes an accelerating force proportional to the difference between the desired  
speed and the present speed, and an influence of other people given by the sum of 
binary person-person interactions which are derived from an exponential potential.
The simplest form depends on the distance only. 
\begin{eqnarray}
F_i = F_{i,acc} + \sum_{i \ne j} F_{i,j}  \qquad \mbox{with} \\
F_{i,acc} = C_1 (v_{i,des} -v_i) ; \\  
F_{i,j}  = C_2\,exp^{ \lambda \, || x_i - x_j || }
\end{eqnarray} 

Forms including the direction of motion
have been developed and are more realistic.
The obstacles are modeled similar to rows of standing people.
While there is little debate on $F_{acc}$, which can easily be tested for 
free movement, the binary interaction is not clear at all, and it 
is not plausible that the combination of binary interactions is additive.
To clarify this issue, we will consider a few simple but extreme setups.

\section{Single lane walking}
The simplest possible situation is that of people walking along a lane, 
and the two extreme cases are a line of people following each other with identical speed
and two people meeting head on in a narrow channel.\\
People following each other is the case of the 1D fundamental diagram (FD)
\cite{SEY05a}. 
This is related to the 2D FD \cite{WEID93,PRED71} via the width of people, 
modified by the zipper effect. This width is speed dependant and not well known, 
so a factor of 1.5 to 3 between pers/s in 1D and pers/(s$\cdot$m) in 2D 
(smaller factors for higher speeds)
is reasonable. With this, the 1D fundamental diagram \cite{SEY05a} is 
compatible with the 2D 
diagrams given. Accuracy cannot be obtained and is not needed here.  
It is clear that the minimal distance obtainable is about 30 cm 
(with cultural differences), while at about 2m distance the full speed 
will be attained so the force should be near zero. The exponential decay of the force 
in eq. 3 is an approximation to this if  $C_2$ and $\lambda$ are appropriate.\\
The first, almost trivial, conclusion to be drawn here is that the force $f_{ij}$ 
cannot be symmetric, otherwise 
the forces from people in front and behind would cancel out and 
the speed would not be density
dependant. Until physical contact occurs, persons behind have 
negligible influence.
A further conclusion is that the combination of the binary 
interactions is not additive. 	 
Actually, assuming $F_{acc}$ given, for a situation with 
identical persons following each other 
at constant speed and distance we have $ F_{i,acc} = 
- \sum_{i \ne j} F_{i,j} $. 
Regardless of which fundamental diagram we
take - there is considerable disagreement between
different measurements - the resulting binary interaction is not plausible.

\subsection{Binary interactions from single lane following}
Keeping in mind that $F_{i,j}=0$ (full speed) for inter-person distances beyond 2m, 
we can directly calculate $F_{i,j}(\Delta x) = - F_{acc}$ for $1$ m $<$ x $<$ $2$ m, 
as there is only one nonzero binary interaction, and the speed corresponding to 
$\Delta x$ and such $F_{acc}$ is known from the FD. \\
For $2/3$ m $<$ x $<$ $1$ m
we have $F_{i,j}(\Delta x) = - F_{acc} - F_{i,j}(2 \Delta x)$, and so on till  \\ 
$F_{i,j}(\Delta x) = - F_{acc} - \sum_{k=2}^7 F_{i,j}(k \Delta x)$ 
for $29$ cm $<$ x $<$ $33$cm.\\
In an 1D experiment with low motivation (therefore not relevant for evacuation), 
we have measured the following 
relation between velocity and distance: 
v$=-0.34+0.945$d for $0.36<$d$<1.314$, \\
v$=0.71+0.146$d for $1.314<$d$<2$,\\
and v$=1$ for d$>2$.
\begin{figure}[htbp]
\epsfig{file=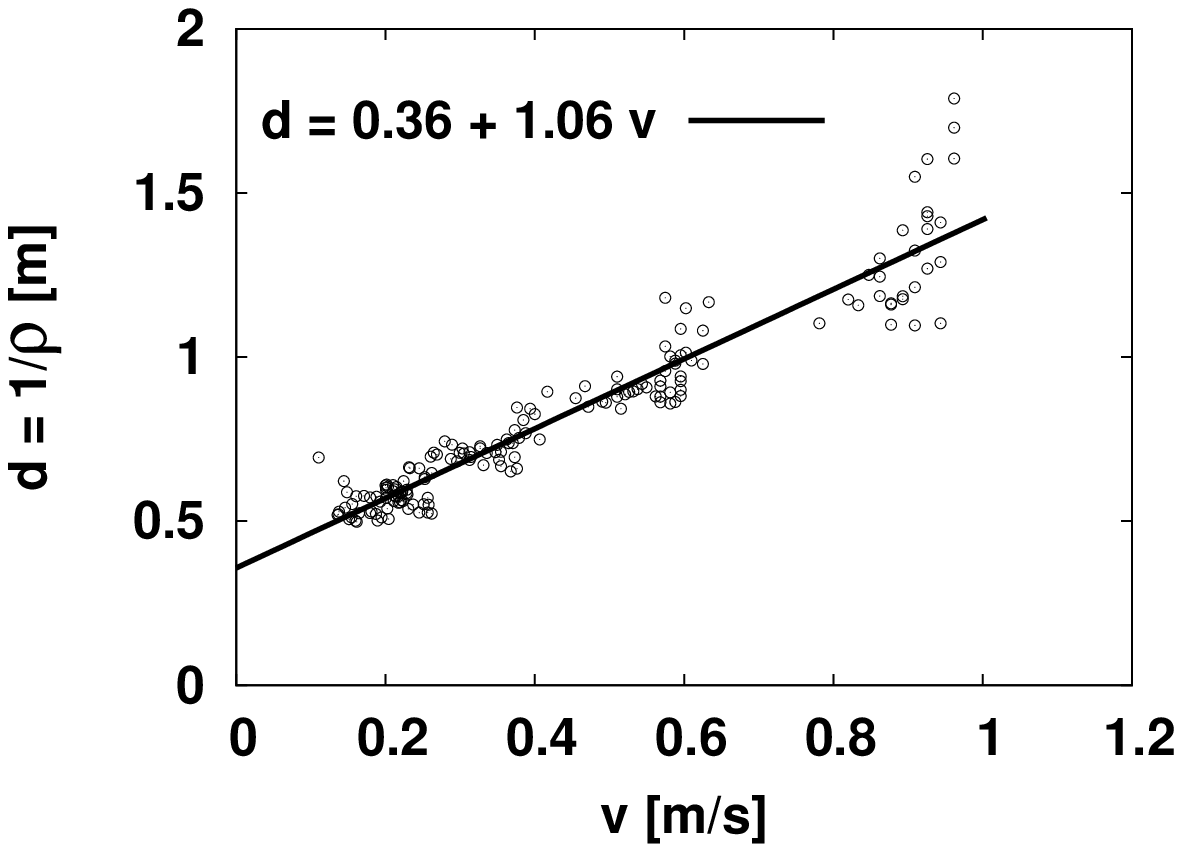,height=4cm,angle=0}
\hspace{0.8cm}
\epsfig{file=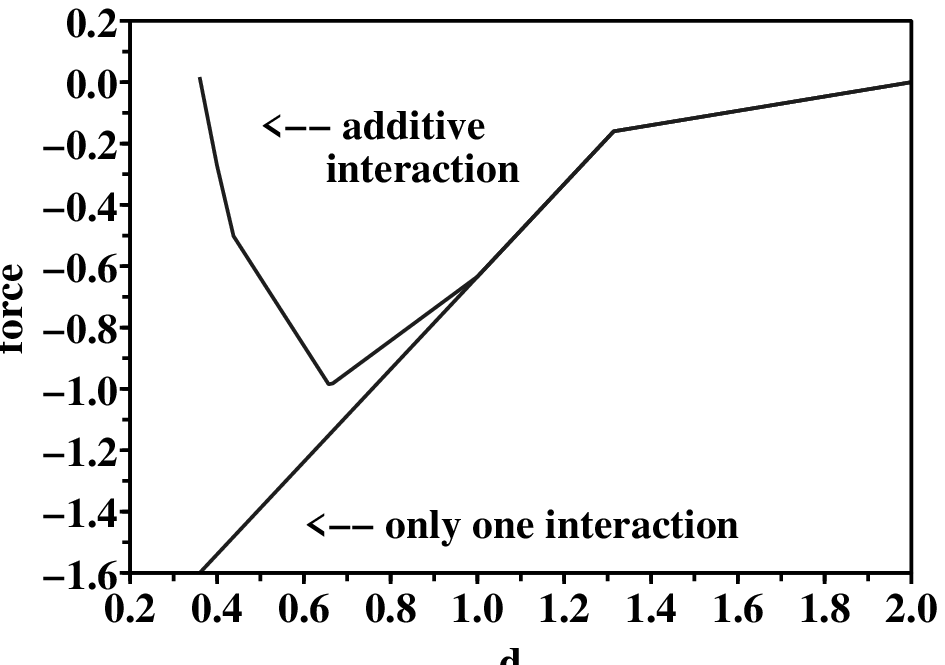,height=4cm,angle=0}
\caption{\rm
  Left:\ Distance versus speed ~~~~
           Right:\ Interaction force calculated from FD}
\label{fig:fit-force}
\end{figure}

Under the additivity assumption, the linear relation fitting the measurements 
gives even attractive binary forces for short distances. Therefore it is reasonable to 
assume that in this situation, there is interaction only with the person 
immediately in front, which gives a reasonable binary interaction. 
This results in the force equation
\begin{eqnarray}
F_{ij} = -2.14 + 1.51 \Delta x, \quad  0.3 < \Delta x < 1.315 \\
F_{ij} = -0.47 + 0.23 \Delta x, \quad  1.315 < \Delta x < 2  
\end{eqnarray}

For more complicated situations, more than one interaction will be relevant, 
but certainly not many.    
For the 1D following,  taking 
only the strongest binary interaction is sufficient.  
More complicated situations may (and do) ask for combining interactions,
but this combination will in general not be additive.

\subsection{Single lane head-on collisions}
Another situation accessible to analytic calculations is the head-on
collision of two persons. In this situation, the force must be strong 
enough to stop both at a reasonable distance from each other.
Further, the speed should come to zero smoothly, especially not show any 
oscillations. Neither the standard social force model nor any force 
extracted from a fundamental diagram by the method described above will 
fulfill these requirements. Further, it is easy to estimate that any
interaction function depending on $\Delta$x only will either be too strong
to give a reasonable speed for walking in file at $\Delta$x=$1.5$m, or 
too weak to handle a head-on collision. Adding a dependence on v alone 
does not help, because any such dependence will be compensated 
by a corresponding change of the dependence on $\Delta$x.
The most simple functional form possible will therefore depend on 
$\Delta$x and on $\Delta$v. A simple and reasonable assumption is that 
a person does not react on the momentary situation, but has some 
foresight and therefore reacts on the extrapolation of the momentary 
situation. More realistic, but more complicated, would be an 
extrapolation out of the recent past to allow for reaction times. 
\cite{HOOG05b} has shown a finite reaction time to be essential, but this
can be taken care of here by shortening the foresight.
With foresight of $1$s, both the binary force from eq.1-3 and the force 
calculated from the FD give reasonable smooth stopping of two persons 
meeting head-on in a narrow corridor.\\
\begin{figure}[ht]
\centerline{\epsfig{file=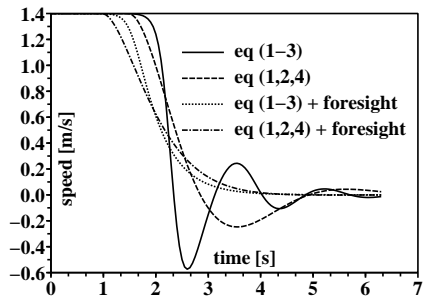,height=4.6cm,angle=0}}
\label{fig:head-on}
\caption{stopping with and without foresight - speed versus time }
\end{figure} 
 
The case of groups in head-on collisions shows that the repulsive force 
is not always given by the closest interaction, and more than a single 
person must be taken into account. The second or third person in a group 
has the closest interaction with somebody going in the same direction, 
while the most relevant interaction is with the first person walking 
in opposite direction. However, the persons in between do not simply 
disappear, so they have to be allotted sufficient space. A reasonable 
procedure is: For each person in front calculate the extrapolated 
distance, reduce it by the space needed by the persons in between, and
calculate the resulting binary force. The repulsive force then is 
the maximum of the binary forces. It should be noted that the 
space needed by persons in between depends on  the speed of the person to 
which the binary interaction is calculated. For a person walking in the 
same direction, this is bigger than for a person with opposite direction,
and it increases with speed.

\section{2D Walking}
The situation has some added complications relative to 
the 1D situation. One is the possibility of partial interference 
without totally blocking the path. These ask e.g. for twisting 
the upper body, which gives some  
delay but no stopping. There is however no data available on how much delay this
would give. The second complication is that people can change directions rapidly. 
Again, exact data is not available, 
but it is easy to test that a single $60^o$ turn can be done within one step without 
much slowing down.    
Because of this possibility, foresight is very restricted in a medium or even 
high density situation where people have reasons to change directions. 
Only in low density situations, where the avoidance behaviour involves only 
a small number of people, a simple foresight model like $1$s extrapolation 
will be beneficial, and more realistic models require knowledge on 
human perception and decision processes that is not available at the moment. \\
\begin{figure}[ht]
\epsfig{file=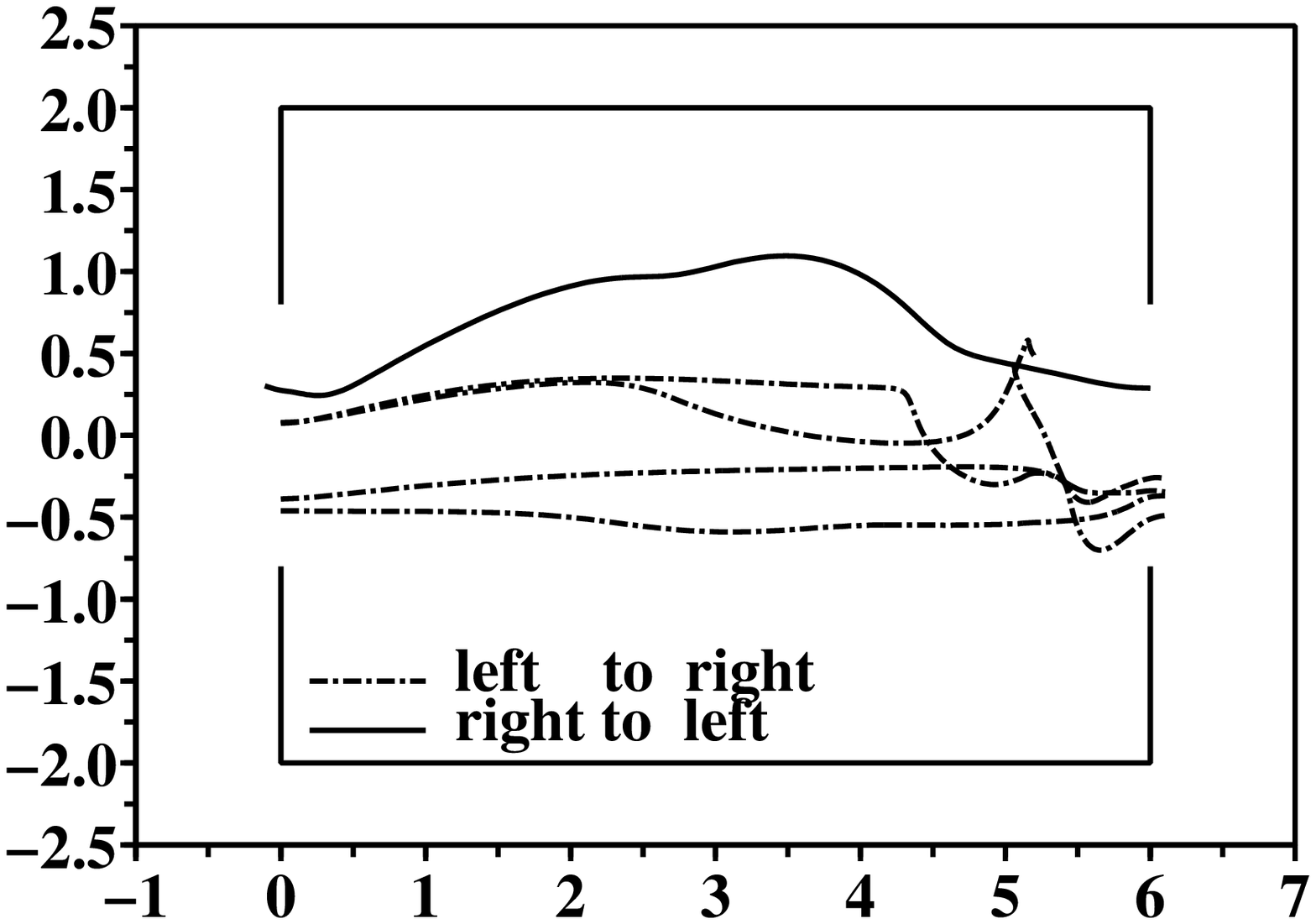,height=4.4cm,angle=0}
\epsfig{file=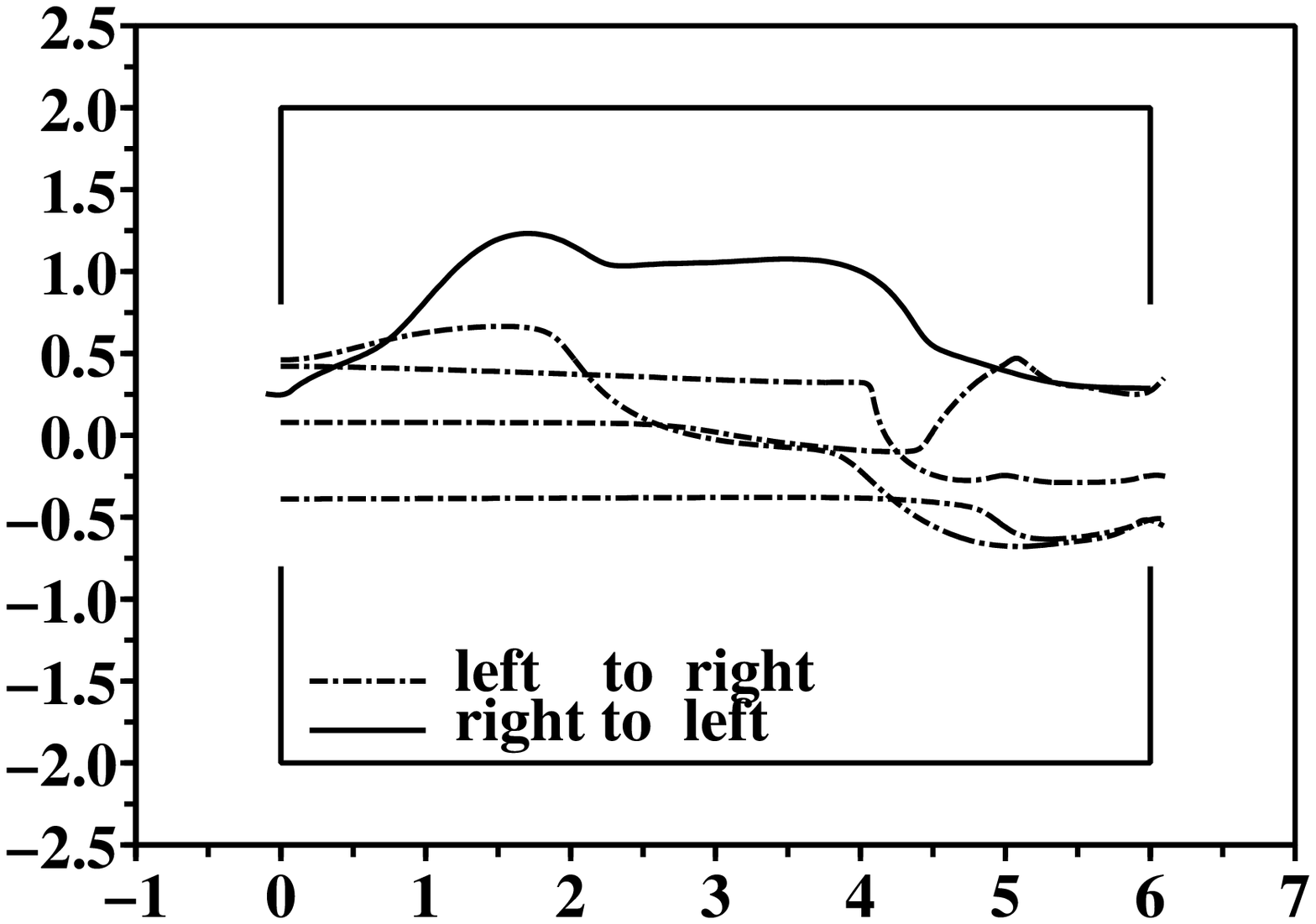 ,height=4.4cm,angle=0}
\caption{Passing a corridor,~~ left: without foresight,~~ right: with foresight  }
\label{fig:Coriidor}
\end{figure}
Still a limited amount of foresight can be helpful. In fig. 3 we compare 
the paths of people passing through a $6\cdot4$ m corridor with 1.4 m 
doors at each side. The walking with foresight is somewhat smoother, and the 
last person out takes $2$s less.   
The difference is small, but enough to look deeper into this subject.

\section{Conclusions}
Simple distance-based repulsive forces are not capable of adequately describing 
microscopic pedestrian behaviour. For 1D head-on collision, the inclusion of a $1$s
foresight resolves this deficiency. In the more interesting low density 
2D situation of meeting in a corridor, the foresight makes the simulation 
a little bit more realistic, but the the collision avoidance still produces 
unrealistic sharp bends and unnecessary slowdowns.

\end{document}